\documentclass[aps, onecolumn, pra, floatfix, superscriptaddress, 10pt]{revtex4-2}
\usepackage[utf8]{inputenc}
\usepackage{graphicx}
\usepackage{bm}
\usepackage{bbold}
\usepackage{physics,amsmath,amssymb,amsfonts,amsthm}
\usepackage{ulem}
\usepackage{csvsimple}
\usepackage{booktabs} 
\usepackage{siunitx}
\usepackage[colorinlistoftodos,prependcaption,textsize=tiny]{todonotes}
\usepackage[colorlinks=true, citecolor=cyan]{hyperref}

\providecommand{\ID}{\ensuremath{\mathbb{1}}}

\begin{document}

\title{Topological Quantum Compilation Using Mixed-Integer 
Programming}
    
	\author{Pavel Ryt\'i\v{r}}
	\affiliation{Department of Computer Science,
		Czech Technical University in Prague, Czech Republic}
	
	\author{Phillip C. Burke}
	\affiliation{School of Physics, University College Dublin, Belfield, Dublin 4, Ireland}
	\affiliation{Centre for Quantum Engineering, Science, and Technology, University College Dublin, Dublin 4, Ireland}
	
	\author{Christos Aravanis}
	\affiliation{Sheffield International College, University of Sheffield,   United Kingdom}
	\affiliation{Department of Computer Science,
		Czech Technical University in Prague, Czech Republic}
			
	\author{Ji\v{r}\'i Vala}
	\affiliation{Department of Physics, Maynooth University, Maynooth, Kildare, Ireland}
    \affiliation{School of Theoretical Physics, Dublin Institute for Advanced Studies, Dublin, Ireland}
    \affiliation{Tyndall National Institute, Cork, Ireland}
    
	\author{Jakub Mare\v{c}ek}
	\affiliation{Department of Computer Science,
		Czech Technical University in Prague, Czech Republic}
	
	\date{\today}

\begin{abstract}
We introduce the Mixed-Integer Quadratically Constrained Quadratic Programming framework for the quantum compilation problem and apply it in the context of topological quantum computing. In this setting, quantum gates are realized by sequences of elementary braids of quasiparticles with exotic fractional statistics in certain two-dimensional topological condensed matter systems, described by effective topological quantum field theories.
We specifically focus on a non-semisimple version of topological field theory, which provides a foundation for an extended theory of Ising anyons and which has recently been shown by Iulianelli et al., Nature Communications {\bf 16}, 6408 (2025), to permit universal quantum computation. While the proofs of this pioneering result are existential in nature, the mixed integer programming provides an approach to explicitly construct quantum gates in topological systems. We demonstrate this by focusing specifically on the entangling controlled-NOT operation, and its local equivalence class, using braiding operations in the non-semisimple Ising system.
This illustrates the utility of the Mixed-Integer Quadratically Constrained Quadratic Programming for topological quantum compilation.
\end{abstract}

\maketitle

\section{Introduction}
Quantum computation \cite{Nielsen_00} can solve an important class of computational problems with a considerable speed-up compared to its classical counterpart. Nevertheless, it faces important challenges due to quantum errors in the computation process. In order for quantum computation to achieve its full quantum power, it has to go beyond the intrinsic limitations of noisy intermediate-scale quantum devices \cite{Preskill_18}. This requires a complex fault-tolerant quantum computing architecture \cite{Shor_96} which is based on quantum error correction with strict error-correction thresholds \cite{Knill_98, Kitaev_03, Aharonov_08}. 
	
Topological quantum computation \cite{Freedman_03, Nayak_08, Wang_10, Pachos_12} promises to bypass these requirements by storing and processing quantum information in topological quantum states. These states are effectively described by topological quantum field theory from which they inherit robustness against errors, such as those due to undesired interactions with stray particles or physical fields. When realized, for example, in a two-dimensional many-body quantum system, these states correspond to ground states with a finite degree of degeneracy, which depends on topological properties of the underlying two-dimensional manifold, such as its genus -- the number of defects, e.g., punctures or the number of quasi-particle excitations. The physical realization of topological quantum computation is an area of intense research activity, with recent examples including, for instance, the work \cite{HW1}, which demonstrated a realization utilizing superconducting circuits, and \cite{HW2}, which demonstrated a realization in ion traps.
	
Point-like excitations, specifically in two-dimensional topological systems, are known as anyons. These excitations are characterized by fractional quantum statistics, which is linked to irreducible representations of the braid group \cite{Kassel_08}. Braiding of these anyons results in a transformation of the quantum state in the anyonic system, which can be either Abelian or non-Abelian. In the non-Abelian case, this transformation corresponds to a unitary rotation of a state vector within the manifold of degenerate ground states of the topological many-body system. The braiding of anyons thus implements quantum computing operations on $n$ quantum bits or qubits, encoded in the ground-state manifold with Hilbert space dimension $d \ge 2^n$. The gate alphabet, native to a topological system, is hence given by a set of elementary braids between non-Abelian anyons. An important feature of anyon theory from the point of view of quantum computation is universality, that is, whether braiding supports universal quantum computing operations from the unitary group $U(2^n)$ over the Hilbert space of $n$ quantum bits.

The compilation problem is a key element in the successful implementation of quantum computation on specific quantum hardware. Its objective is to convert a quantum algorithm or protocol into a form that can then be executed on a quantum computing device. Quantum hardware is typically characterized by various constraints, such as qubit connectivity and native gate alphabet. The goal of topological quantum compiling is to design a specific sequence of the elementary braids in order to implement a desired quantum computing gate, which unitarily transforms a qubit state encoded into a multiply degenerate ground state of a given topological system. 

The seminal works on topological quantum compilation~\cite{Bonesteel_BraidTQC_PRL2005, Simon_06, Hormozi_TQC_PRB2007} primarily use a three-anyon encoding of a logical qubit to realize both single-qubit and two-qubit operations via braiding or `weaving' of Fibonacci anyons. In more recent work, Carnahan et al.~\cite{Carnahan_16} developed a systematic approach to the generation of two-qubit anyon braids based on an iterative procedure, which was proposed earlier by Reichardt~\cite{Reichardt_12}. Kliuchnikov et al.~\cite{Kliuchnikov_14} developed an asymptotically optimal probabilistic polynomial algorithm to generate single-qubit gates exactly and also to produce depth-optimal approximations of two-qubit gates in the Fibonacci anyon model. Relevant to the topological quantum compilation problem is also the work on topological hashing by Burrello et al.~\cite{Burrello_10} and a recent application of reinforcement learning for quantum compilation~\cite{Zhang_20, Chen_2024}.

The implementation of gates in topological quantum computing typically requires elementary braiding operations acting on a Hilbert space of a larger dimension than that of two qubits. The problem of leakage in relation to topological quantum compilation, particularly in the context of the implementation of two-qubit gates, was recognized early on by Bonesteel et al.~\cite{Bonesteel_BraidTQC_PRL2005}. This problem was studied later by Xu and Wan~\cite{Xu_08}, and also by Ainsworth and Slingerland~\cite{Ainsworth_11}, who focused on the leakage problem in connection with topological qubit design and showed that the requirement of leakage-free design is too restrictive and can only be realized in non-universal Ising-like anyon models. A large family of leakage-free Fibonacci braiding gates were studied by Cui et al.~\cite{Cui_2Qubit_JoPA2019}, who proved that they were all non-entangling. Moreover, brute-force numerical searches failed to identify any leakage-free implementations of entangling operations using braids with a word length of up to seven. 

Recently, Iulianelli et al.~\cite{iulianelliUniversalQuantumComputation2024} proposed a new framework for topological quantum computation based on non-semisimple extensions of $2+1$-dimensional topological quantum field theories. These enhanced models go beyond conventional Ising anyons, which are associated with the $\nu = \frac{5}{2}$ fractional quantum Hall state~\cite{stormer1999nobel} but which are not universal for quantum computation via braiding alone. In the non-semisimple theory, the Ising framework is extended by a single new anyon which enables universal quantum computation solely through braiding, thus lifting the Ising anyon system to the universal topological quantum computation.

In this work, we provide a Mixed Integer Quadratically Constrained Quadratic Programming (MIQCQP) formulation for the gate compilation problem. As a case study, we demonstrate our approach on a system of anyons that emerge in a non-semisimple version of the Ising topological quantum field theory ~\cite{iulianelliUniversalQuantumComputation2024}.
We specifically aim to compile two-qubit entangling gates, such as the controlled-NOT (CNOT) gate and its local equivalence class, as well as other perfect entanglers. We employ the geometric theory of two-qubit operations, developed by Zhang et al.~\cite{ZhangValaBirgitta_GeometricTheory_PRA2003} in the context of quantum control. The geometric theory stems from the Cartan decomposition of the Lie group $SU(4)$ and the Makhlin local invariants~\cite{Makhlin_Invariants_QIP2002}, which provides a representation of two-qubit operations in terms of their local equivalence classes. In our context, a local equivalence class is defined as the set of two-qubit gates that are equivalent up to single-qubit transformations, and thus correspond to the co-set of $SU(4)$ by the subgroup $SU(2) \otimes SU(2)$. We will use the gate alphabet proposed in the paper~\cite{iulianelliUniversalQuantumComputation2024}.

The recent work by Burke et al.~\cite{Burke_24} significantly extended the calculations by Cui et al.~\cite{Cui_2Qubit_JoPA2019} to larger braiding combinations and also employed the theory of local invariants and equivalence classes of two-qubit operations. The local invariants were previously used by DiVincenzo et al.~\cite{DiVincenzo_00} to find a local equivalent of the CNOT gate in a system with the Heisenberg Hamiltonian in the context of encoded universality. M\"{u}ller et al.~\cite{Muller_11} have combined optimal quantum control with the geometric theory of two-qubit operations to derive an optimization algorithm that determines the best two-qubit entangling gate in the system of trapped polar molecules and neutral atoms. This work was later followed by a generalization of the optimal control objective to include an arbitrary perfect entangler~\cite{Watts_15, Goerz_15}. In all these cases, the use of geometric theory, and the Makhlin local invariants in particular, led to a significant enlargement of the optimization target, its dimension, and size, which considerably improved both the accuracy and convergence of the optimization process. 

The paper is organized as follows. In Section \ref{sec_MIQCQP}, we present the mixed integer programming formulation of quantum compilation. Section \ref{sec_isinganyons} briefly reviews the main features of the system of Ising anyons in a non-semisimple topological quantum field theory. Section \ref{sec_results} presents the experimental results of topological quantum compilation in this system. Finally, the conclusion follows in Section \ref{sec_conclusions}.


\section{Mixed Integer Programming Formulation}\label{sec_MIQCQP}

The goal of this section is to formulate the quantum compilation problem in the MIQCQP framework~\cite{tawarmalaniConvexificationGlobalOptimization2002, boydConvexOptimization2023a, burerNonconvexMixedintegerNonlinear2012}.  This optimization technique is particularly suitable for optimization problems in which some of the variables are discrete. The main advantage of this method, from our perspective, is that it provides global optimality guarantees. This, in turn, prevents the braid sequences resulting from the compilation of entangling quantum operations from being unnecessarily long. Another advantage of MIQCQP is that it has been continuously developed and extensively utilized, for example, in optimizing logistics problems. Hence, there are many mature MIQCQP solvers with global optimality guarantees available. This makes the method well understood, including its computational complexity. MIQCQP is, in general, NP-hard but can be more favorable in specific instances. Moreover, its widespread usage carries over to the quantum compilation problem when formulated in the MIQCQP framework. 
In the following subsections, we provide a brief outline of the quantum compilation problem and then give a detailed description of the MIQCQP formulation in the quantum compilation context.
\subsection{Quantum Compilation Problem}\label{sec_circuitcomp}

The exact (i.e., not locally equivalent) quantum compilation problem aims to find a sequence of gates $I$ from a specified basic alphabet of elementary operations or gates that equals the desired target operation $T$. The quantum compilation problem can be formally defined as follows:
Let $\{\rho_1, \dots, \rho_m\}\in \mathbb{C}^{n\times n}$ be the matrices representing the basic gate alphabet, and $T\in\mathbb{C}^{n_\text{C}\times n_\text{C}}$ be a target gate, with $n_\text{C}\leq n$; $n_\text{C},n\in\mathbb{N}$. Here, $n_\text{C}$ is the dimension of the computational subspace, $V_{\text{C}}$.
We define the unitary matrix $P$ as the resultant operation corresponding to a sequence $I$ as
\begin{equation}
    P = \prod_{i\in I}\rho_i.
\end{equation}
A unitary operation $P$ is said to be leakage-free if it maps the computational subspace $V_\text{C}$ onto itself. Thus, a leakage-free operation must have the block-diagonal structure
\begin{equation}
    P=X\oplus U,
\end{equation}
where $U$ is a unitary matrix acting on the computational subspace $V_\text{C}$ with dimension $n_\text{C}$, and $X$ is 
a unitary matrix acting on the non-computational subspace $V_\text{NC}$ of dimension $n_\text{NC}$. The total dimension of the Hilbert space is thus $n=n_\text{C}+n_\text{NC}$. 

From a practical point of view, it is useful to relax the notion of exact quantum compiling to its approximate form, in which one or several objectives are achieved within a desired accuracy.
This may include the approximation of the target gate $T$, or approximate unitarity (or approximate leakage-free realization) in the computational subspace $V_\text{C}$ within an accuracy given by some suitable distance metric. We will specify these in a concrete example below.

\subsection{Program Variables}

We shall assume that the circuit has a fixed depth $d$. If we do not know the exact depth in advance (which is the case for our compilation problem), we can include an identity matrix in the basic alphabet, permitting us to search for a circuit of maximum depth $d$.
With $d$ the fixed depth of the circuit, the compiled circuit is constructed in $d$ steps.
For $t=0,\dots, d$; we define the complex variables $[Y_{t}]_{i,j}\in\mathbb{C}$ for $i,j=1,\dots,n$. These variables constitute the matrix $Y_t$, which encodes the circuit obtained as a product of $t$ gates from the basic alphabet, i.e.
\begin{equation}
    Y_t=\prod_{i\in I_t}\rho_i,
\end{equation}
where $I_t$ is a prefix of $I$ with a length of $t$.

At each step $t=1,\dots,d$, we have to decide which gate from the basic alphabet of size $m$ should be applied. This decision is encoded as a binary decision variable $x_{t,i}\in\{0,1\}$, for $t=1,\dots,d$ and $i\in 1,\dots,m$.
Here, $x_{t,i}=1$ if, and only if, gate $\rho_i$ is applied at step $t$, in which case, $x_{t,j}=0 \ , \ \forall j \neq i$.

\subsection{Program Constraints}
First, we must define the initial state constraints. These constraints say that the empty product of gates from the basic alphabet equals the $n\times n$ identity matrix.
\begin{equation}
    Y_{0} = \mathbb{1}_n.
\end{equation}
In each step $t=1,\dots,d$, we multiply the product $Y_{t-1}$, constructed in the previous step $t-1$, by the chosen gate from the basic alphabet. The relation between $Y_t$ and $Y_{t-1}$ is encoded in the following matrix constraint.
\begin{equation}\label{eq:states_multiplications}
    Y_t=\left(\sum_{i=1}^m x_{t,i}\rho_i\right)Y_{t-1},
\end{equation}
where $x_{t,i}$ are the binary decision variables, selecting which gate from the basic alphabet is applied at step $t$. We reiterate that exactly one gate can be applied in each step. This is enforced by the following constraint:
\begin{equation}\label{eq:gate_selection}
    \sum_{i=1}^mx_{t,i}=1,\quad \text{for every step }t=1,\dots,d.
\end{equation}

Finally, we introduce constraints that restrict leakage to the non-computational space of the final product $Y_d$.
We wish to enforce that $Y_d=X\oplus U$, where $U$ is the submatrix that corresponds to the computational space. By setting the elements of $Y_d$ outside the blocks $X$ and $U$ to zero. Then, the matrix $U$ has to be unitary as $Y_d$ is unitary, and thus there is no leakage to the non-computational space.
We recall that an $m\times m$ matrix $U$ is defined to be unitary if $U^\dagger U =UU^\dagger=\mathbb{1}$.

\subsection{Program Objective Function}
We now need to define the objective function for the problem.
Due to numerical errors and inaccuracies, it is unlikely that we could find a product of basic alphabet gates that is exactly equal to the target gate. Therefore, we define the distance $J$ between the target gate and the computational part of the computed product $Y_d$.
The distance $J$ is defined as the square of the Frobenius norm of the difference between $T$ and $U$,
\begin{equation}\label{eq:objective}
    J(T,U)=\lVert T-U\rVert^2_F=\sum_{k,\ell=1}^{n}\left(T_{k,\ell}-U_{k,\ell}\right)^2.
\end{equation}
Our goal will be to minimize this distance, i.e., this is our objective function. 
\subsection{MIQCQP Problem Formulation}
The mixed integer quadratic program for the gate compilation problem is defined as follows. 
Let $Y^\text{C}_d$ denote the computational part of the matrix $Y_d$ at the final step $d$.

Our objective is to minimize 
\begin{equation}
    J(T, Y^\text{C}_d)=\sum_{k,\ell=n_\text{NC}+1}^{n}\left(T_{k,\ell}-[Y_{d}]_{k,\ell}\right)^2.
\end{equation}
The optimization procedure is subject to the following constraints:
\begin{alignat}{2}  
    & \text{(i) }[Y_t]_{k,\ell} = \sum_{p=1}^{n} \left( \sum_{i=1}^{m} x_{t,i}[\rho_{i}]_{k,p}[Y_{t-1}]_{p,\ell}\right), \,\,\, \text{ for } k,\ell=1,\dots, n, \text{ and } t=1,\dots,d\label{eq:states_multiplication_entrywise}\\
    & \text{(ii) }\sum_{i=1}^{m}x_{t,i} = 1, \text{ for each } t=1,\dots,d,\\
    & \text{(iii) }x_{t,i} \in \{0,1\}, \text{ for } t=1,\dots,d, \text{ and } i=1,\dots m,\\
    & \text{(iv) } [Y_0]_{k,\ell} = \delta_{k,l}, \text{ for } k,l=n_\text{NC} + 1,\dots, n,\\
    & \text{(v) }[Y_d]_{k,\ell} = 0, \text{ for } k = 1,\dots, n_\text{NC};\,\, \ell=n_\text{NC}+1,\dots, n,\\
    & \text{(vi) }[Y_d]_{k,\ell} = 0, \text{ for } k = n_\text{NC}+1,\dots, n;\,\, \ell=1,\dots, n_\text{NC},\\
    & \text{(vii) }[Y_t]_{k,\ell}\in\mathbb{C}, \text{ for } t=0,\dots,d, \text{ and } k,l=1,\dots n.
\end{alignat}

The constraints (i) correspond element-wise to Eq.~\eqref{eq:states_multiplications}, while the constraints (ii) correspond to Eq.~\eqref{eq:gate_selection}. 
The constraints (iii) enforce integrality on the gate selection variables, and (iv) set the initial empty product to be the $n\times n$ identity matrix. The constraints (v) and (vi) enforce the final product $Y_d$ to have the form $X\oplus Y^\text{C}_d$, where $X$ acts on the non-computational space and $Y^\text{C}_d$ on the computational space. These constraints also ensure that $Y^\text{C}_d$ to be unitary, since $Y_d$ is unitary. Finally, the constraints (vii) specify that the product variables are complex numbers.


\subsection{MIQCQP Problem Relaxation and Linearization}
The performance of solvers on MIQCQP programs is usually more time-consuming than on mixed integer linear programs (MILP). This is due to three factors: the root relaxation of an MIQCQP is being solved with an interior-point method that cannot exploit as much sparsity as in the MILP case, and may need a cross-over routine. Relaxations solved subsequently in the branch-and-bound tree are also more difficult to solve, as the dual simplex commonly used in MILP does not generalize easily. Finally, there are stronger cuts applicable to MILP than MIQCQP, which can increase the number of nodes visited in the branch-and-bound process. In proof complexity \cite{hrubevs2017random, Pfetsch2024}, one could provide lower bounds for the final aspect. 

The constraints in Eq.~\eqref{eq:states_multiplications}
and the objective function Eq.~\eqref{eq:objective} are non-linear. This can negatively affect the performance of the solver. The McCormick linearization procedure~\cite{mccormickComputabilityGlobalSolutions1976} replaces a bilinear constraint $z=ab$, where $a$ is a binary variable and $b$ is a continuous variable, by the following set of linear constraints. Let $b_\text{min}$ and $b_\text{max}$ be the minimal and maximal allowed values of $b$, respectively.
\begin{equation}\label{eq:mccormick}
\begin{aligned}
    z &\geq b_\text{min}a, \\
    z &\geq b + b_\text{max}a + b_\text{min}, \\
    z &\leq b_\text{max}a, \\
    z &\leq b + b_\text{min}a+b_\text{max}.
\end{aligned}
\end{equation} 

This linearization is exact; replacing the bilinear constraint $z=ab$ with the four linear constraints~\eqref{eq:mccormick} yields the exact same set of solutions.

Using this McCormick linearization, all constraints in Eq.~\eqref{eq:states_multiplication_entrywise} can be replaced by a set of linear constraints. The Equation~\eqref{eq:states_multiplication_entrywise} is:
\begin{equation}
    [Y_t]_{k,\ell} = \sum_{p=1}^{n} \left( \sum_{i=1}^{m} x_{t,i} [\rho_{i}]_{k,p}[Y_{t-1}]_{p,l} \right), \label{eq:states_multiplication_entrywise_remind}
\end{equation}
The right-hand side of Eq.~\eqref{eq:states_multiplication_entrywise_remind} is the sum of the bilinear expressions 
\begin{equation}\label{eq:z_expr}
    x_{t,i}[\rho_{i}]_{k,p}[Y_{t-1}]_{p,l},
\end{equation}
where $[\rho_{i}]_{k,p}$ is a constant.
We apply McCormick linearization to transform the bilinear constraint Eq.~\eqref{eq:states_multiplication_entrywise_remind} into a set of linear constraints without worsening the solution quality, as follows. For each $i=1,\dots,m$; $k=1,\dots,n$; $l=1,\dots,n$; $p=1,\dots,n$; and $t=1,\dots,d$, we introduce a new variable, $z_{t,k,l}^{p,i}$, which is constrained to take the same value as the expression in Eq.~\eqref{eq:z_expr}. For each $z_{t,k,l}^{p,i}$ we then add the following four constraints:
\begin{equation}\label{eq:mccormick_applied}
\begin{aligned}
    z_{t,k,l}^{p,i} &\geq -\lvert[\rho_{i}]_{k,p}\rvert x_{t,i}, \\
    z_{t,k,l}^{p,i} &\geq [\rho_{i}]_{k,p} [Y_{t-1}]_{p,l} + \lvert[\rho_{i}]_{k,p}\rvert x_{t,i}  -\lvert[\rho_{i}]_{k,p}\rvert, \\
    z_{t,k,l}^{p,i} &\leq \lvert[\rho_{i}]_{k,p}\rvert x_{t,i}, \\
    z_{t,k,l}^{p,i} &\leq [\rho_{i}]_{k,p} [Y_{t-1}]_{p,l}  -\lvert[\rho_{i}]_{k,p}\rvert x_{t,i}+\lvert[\rho_{i}]_{k,p}\rvert.
\end{aligned}
\end{equation} 
Finally, the constraint of Eq.~\eqref{eq:states_multiplication_entrywise_remind} is replaced by the following linear constraint.
\begin{equation}
    [Y_t]_{k,\ell} = \sum_{p=1}^{n} \left( \sum_{i=1}^{m} z_{t,k,l}^{p,i} \right).\label{eq:states_multiplication_entrywise_substituted}
\end{equation}

\subsection{Local Equivalence Classes}

We will see later that finding a specific operation, such as the CNOT gate, is not only a formidable task but also a frequently unsuccessful one. In the case of two-qubit operations, this amounts to searching for a specific point in the manifold of two-qubit operations from the Lie group $SU(4)$, which has fifteen dimensions. 

Since we are primarily interested in operations that can change entanglement between two quantum bits, we can replace a target gate $T$ as the compilation objective by its local equivalence class $[T]$, that is, by all unitary operations that are equivalent to $T$ up to arbitrary single-qubit transformations~\cite{ZhangValaBirgitta_GeometricTheory_PRA2003}. This significantly increases the size of an optimization target from a point in the fifteen-dimensional manifold to its twelve-dimensional subspace. In other words, an optimization reduces to a search in the three-dimensional space of local equivalence classes~\cite{Muller_11}.

Particularly useful instruments for defining local equivalence classes are the local invariants defined by Makhlin~\cite{makhlinNonlocalPropertiesTwoqubit2002} and further studied in the context of quantum control by Zhang et al.~\cite{ZhangValaBirgitta_GeometricTheory_PRA2003}, where more details about their construction can be found.

A two-qubit unitary operator $U \in SU(4)$ has the Cartan decomposition 
\begin{equation}\label{eq:u}
    U = k_1Ak_2
\end{equation}
where $k_1$ and $k_2$ are from the single-qubit subgroup $SU(2) \otimes SU(2)$ of $SU(4)$. Importantly, $A$, which is generated by the Cartan subalgebra of the Lie algebra $SU(4)$, encodes all the non-local or entangling properties of the unitary operation.

We can define the local invariants as follows: 
We first transform the matrix $U$ into the Bell basis 
\begin{equation}\label{eq:ub}
    U_B = \mathcal{Q}^{\dagger} U \mathcal{Q}.
\end{equation}
using the matrix 
\begin{equation}
		\mathcal{Q} = \frac{1}{\sqrt{2}}
		\begin{pmatrix}
			1 & 0 & 0 & i \\
			0 & i & 1 & 0 \\
			0 & i & -1 & 0 \\
			1 & 0 & 0 & -i
		\end{pmatrix}.
	\end{equation}
This transformation makes the single qubit terms of the Cartan decomposition $k_1$ and $k_2$ elements of the special orthogonal group $SO(4)$. The non-local matrix $A$ transforms into a diagonal form. Then, calculating the Makhlin matrix
\begin{equation}\label{eq:mu}
    m_U = U_B^{T}~ U_B,
\end{equation}
eliminates one of the local components, as the product $O^TO$ is the identity for $O \in SO(4)$. The second of the local operations can be shown to diagonalize the Makhlin matrix $m_U$. From the characteristic equation and the factorization of an arbitrary unitary matrix from the group $U(4)$ into a product of the global phase $U(1)$ and $SU(4)$, we can derive the local invariants for an arbitrary two-qubit unitary matrix in the following form:
\begin{equation}\label{eq:g}
g_1 =  \operatorname{Re}\left\{\frac{\operatorname{tr}^2(m_U)}{16 \det(U)}\right\}, \quad
g_2 =  \operatorname{Im}\left\{\frac{\operatorname{tr}^2(m_U)}{16 \det(U)}\right\}, \quad
g_3 = \frac{\operatorname{tr}^2(m_U) - \operatorname{tr}(m_U^2)}{4 \det(U)}.
\end{equation}
Instead of compiling exactly a target gate, we could instead attempt to compile a gate that belongs to the local equivalence class of the target gate ~\cite{Muller_11}. 
A suitable "distance" measure to the CNOT local equivalence class can be defined using these invariants, for example, as in~\cite{Muller_11}
\begin{equation}
    D_\text{CNOT}(U)=g_1^2+g_2^2+(g_3-1)^2.
    \label{eq_CNOT_dist}
\end{equation}
In some applications, we may be interested in generating an arbitrary perfect entangler, that is, an operation capable of creating a maximally entangled state out of some product state.
This problem can be solved by using the cost function developed in~\cite{wattsOptimizingArbitraryPerfect2015}.
This cost function can be defined in the following form:
\begin{equation}
    D_\text{PE}(U) = \left(g_1 - g_3 \sqrt{g_1^2 + g_2^2}\right)^2.
    \label{eq_PE_dist}
\end{equation}

\subsection{Distance to the Local Equivalence Class MIQCQP Problem Formulation}
In this section, we elaborate on how to implement the cost function from the previous section as an MIQCQP program. The encoding of equations~\eqref{eq:ub} and \eqref{eq:mu} is straightforward and analogous to the encoding of Eq.~\eqref{eq:states_multiplications}.
The division $c=a/b$ can be implemented as a bilinear constraint $a = bc$. 
The determinant can be calculated using a recursive Laplace (cofactor) expansion~\cite[Page 752, Section VI.23]{gowersPrincetonCompanionMathematics2008} that leads to bilinear constraints.
The constraints are as follows: For a given matrix $A$, we define a variable $z_A$ that is constrained to have the value of $\det(A)$.
If $A$ is a $2\times 2$ matrix, then the following bilinear constraint ensures that $z_A=\det(A)$,
\begin{equation}
    z_A = A_{11}A_{22}-A_{12}A_{21}.
\end{equation}
If $A$ is an $n\times n$ matrix, where $n>2$. Then
\begin{equation}\label{eq:det_exp}
    z_A=\sum_{j=1}^{n} (-1)^{j} A_{1j} z_{A,1j},
\end{equation}
where variable $z_{A,1j}$ is constrained to have a value of the determinant of matrix $A$ with the first row and j-th column removed, which can be achieved by recursively adding the constraint~\eqref{eq:det_exp} for the submatrices of $A$.

For the compilation of a gate in the CNOT local equivalence class, we use the following cost function:
\begin{equation}
    J(Y^\text{C}_d)=D_\text{CNOT}(Y^\text{C}_d).
\end{equation}
While for the compilation of a general perfect entangler gate, we use the following cost function:
\begin{equation}
    J(Y^\text{C}_d)=D_\text{PE}(Y^\text{C}_d).
\end{equation}

Note that we do not replace any of the constraints used for computing either  $D_\text{CNOT}(Y^\text{C}_d)$ or $D_\text{PE}(Y^\text{C}_d)$ using McCormick linearization, as this linearization is not exact for bilinear forms where both variables are continuous.

\subsection{Solving the MIQCQP Programs}

The above MIQCQP programs can be implemented using various free or commercial solvers such as CPLEX, SCIP, or Gurobi. In our work, we implemented the MIQCQP programs in Gurobi~\cite{gurobi}.
The Gurobi solver for mixed-integer quadratic programs is based on a branch-and-bound-and-cut framework~\cite{mitchellBranchandCutAlgorithmsCombinatorial2002a}, and it employs the simplex method~\cite{dantzig1947simplex}, as well as interior-point and barrier methods~\cite{karmarkar1984new, nocedal2006numerical}, to solve the linear and quadratic programming relaxations.

Note that MIQCQP solvers typically do not operate directly on complex arithmetic. This limitation can be overcome by decomposing all complex expressions into their real and imaginary components and formulating the constraints and objective function in terms of these components.

\section{Ising anyons from a non-semisimple Topological Quantum Field Theory}\label{sec_isinganyons}
In order to experimentally evaluate the proposed method, we will use a framework for topological quantum computation using newly discovered nonsemisimple topological quantum field theory in 2+1 dimensions \cite{iulianelliUniversalQuantumComputation2024}.

The basic gate alphabet is defined as follows. Let $\alpha\in\mathbb{R}$ be a parameter. Let $q$ be an eighth root of unity, so $q$ can be expressed as $e^{\frac{2\pi i k}{8}} $ for some $ k \in\{0,...,7\}$. In our experiments, we will set $q=e^{\frac{2\pi i}{8}}$.

First, the following real coefficients are defined:

\begin{equation}
B_{\alpha+1} := \frac{\sqrt{2}}{-1 + \cot \frac{\pi (\alpha+1)}{4}}, \quad
B_{\alpha-1} := \frac{\sqrt{2}}{-1 + \cot \frac{\pi \alpha}{4}}.
\end{equation}
For braiding single qubits the matrices are:

\begin{equation}
b_1^2 = -q \left( b_1^{\alpha \sigma \sigma} \right)^2 =
\begin{pmatrix}
    q^\alpha & 0 \\
    0 & q^{-\alpha}
\end{pmatrix},
\end{equation}

\begin{equation}
b_2 = q^{-\frac{3}{2}} b_2^{\alpha \sigma \sigma} = q^{-1}
\begin{pmatrix}
    \frac{1 + q^2}{1 - q^{2\alpha}} & q^{-1} \frac{\sqrt{B_{\alpha+1}}}{\sqrt{B_{\alpha-1}}} \\
    q^{-1} \frac{\sqrt{B_{\alpha+1}}}{\sqrt{B_{\alpha-1}}} & \frac{1 + q^2}{1 - q^{-2\alpha}}
\end{pmatrix}
\end{equation}

Let $\ID_2$ be the $2\times 2$ identity matrix. The operations on the first qubit of the 2-qubit system are performed by the following two matrices:
\begin{equation}
b_1^2 = \left( b_1^{\alpha \sigma \sigma} \right)^2 \otimes \ID_2 \oplus \left( b_1^{\alpha \sigma \sigma} \right)^2,
\end{equation}

\begin{equation}
b_2 = b_2^{\alpha \sigma \sigma} \otimes \ID_2 \oplus 
\begin{pmatrix}
    q^{\frac{1}{2}} \ID_2
\end{pmatrix}.
\end{equation}

The operations on the second qubit of the 2-qubit system are performed by the following two matrices:

\begin{equation}
J_4 = \text{Id} \otimes \left( b_1^{\alpha \sigma \sigma} \right)^2 \oplus
\begin{pmatrix}
    q^{1-\alpha} & 0 \\
    0 & q^{1+\alpha}
\end{pmatrix},
\end{equation}

\begin{equation}
b_4 = \ID_2 \otimes b_2^{\alpha \sigma \sigma} \oplus 
\begin{pmatrix}
    q^{\frac{1}{2}} \ID_2
\end{pmatrix}.
\end{equation}

Finally, an entangling gate with very low leakage into the non-computational space is constructed in~\cite{iulianelliUniversalQuantumComputation2024} for $\alpha=2.4$ using an iterative method by Reichardt~\cite{reichardtSystematicDistillationComposite2012}.
Since the leakage term of the constructed entangling gate has a very low leakage term, approximately $2.565 \times 10^{-14}$, we will assume that the gate is leakage free, in order to simplify our calculations.

The entangling gate is a controlled phase shift and is given by:
\begin{equation} \label{eq_entanglingMatrix}
    \text{CPHASE} = \begin{bmatrix}
    1 & 0 & 0 & 0 \\
    0 & 1 & 0 & 0 \\
    0 & 0 & e^{-1.772i} & 0 \\
    0 & 0 & 0 & e^{-1.682i}
    \end{bmatrix}\oplus V, 
\end{equation}

where $V$ is a unitary $2\times2$ matrix.

Finally, the basic gate alphabet, denoted by $\text{NON-SEMI}$, is equal to $\{b_1^2, b_2, J_4, b_4, \text{CPHASE}\}$.

\section{Experimental Results}\label{sec_results}
The experiments are divided into three subsections.
In the first subsection, we attempt to compile the CNOT gate.
In the second subsection, we search for gates that are as close as possible to the CNOT local equivalence class. In the third subsection, we will broaden our search to the set of perfect entangling gates.

We evaluate our experiments via two metrics: The cost function $J$ and the Hilbert-Schmidt distance between to the desired gat.
The (normalized) Hilbert-Schmidt distance is defined as follows:
\begin{equation}\label{eq_d2}
    d_2(A,B) = \left\| \frac{A}{\|A\|_2} - \frac{B}{\|B\|_2}\right\|_2,
\end{equation}
where $\|.\|_2$ is the Hilbert-Schmidt ($p=2$) norm, and each matrix is normalized by its norm. This normalization ensures the distances are bounded between $0$ and $2$, providing a sense of scale. We note that this measure can return a zero value for proportional matrices due to this normalization.

The experiments were run on a computing cluster with AMD EPYC 7543 CPUs. Each experiment was allocated 4 CPU cores and 32 GB of memory. The time limit for the Gurobi solver was set to 23 hours, after which we recorded the best solution found, which was not necessarily optimal.

\subsection{CNOT gate compilation}
A natural candidate for the demonstration of our method is the CNOT gate. This gate operates on two qubits and can change the degree of entanglement between them. Since it can also generate a maximally entangled state of two qubits from some initial non-entangled state, it is also an example of a perfect entangler. In the first set of experiments, we employ the MIQCQP to find a realization of the CNOT gate as a sequence of the operations from the entire gate alphabet $\text{NON-SEMI}$. In accord with~\cite{iulianelliUniversalQuantumComputation2024}, we use as a part of the gate alphabet the $\text{CPHASE}$ operation, which is defined in Eq.~\eqref{eq_entanglingMatrix} above, and which is obtained for the value of the parameter $\alpha=2.4$. The depth of the circuit, consisting of the elements of the gate alphabet, ranged from $10$ to $200$ in the optimization process. 

The results are shown in Table~\ref{tab:cnot24_gate_compilation} which lists the best outcomes of this type of experiment. We observe that this quantum compilation process has manifestly failed to find CNOT as a circuit of depth to most 200. The native alphabet from $\text{NON-SEMI}$ is unable to reach this gate in 200 steps. CNOT, like any element of the Lie group $U(4)$, factorizes into a trivial $U(1)$ factor and an element of $SU(4)$. To find a specific operation in $SU(4)$ is then equivalent to finding a point in the manifold of fifteen dimensions. A similar situation, albeit in a different anyonic system, was encountered in ~\cite{Burke_24}. It seems plausible that $\text{NON-SEMI}$, with its limited number of elements, cannot reach the CNOT target in a reasonable number of steps, even though this non-semisimple anyon system is universal for quantum computation~\cite{iulianelliUniversalQuantumComputation2024}.

\begin{table}[]
    \centering
    \begin{tabular}{rrrp{12cm}}
\toprule
Depth & $J(\cdot)$ & $d_2(\cdot,\text{CNOT})$ & Program \\
\midrule

150 & 0.163624 & 0.202252 & 44444444444440444444444444444444444443444444444444444444404444444440444440 44444444444444444444444444444444444444444244444444444444444444444044444444
23 \\
161 & 0.353581 & 0.297313 & 44444004444444444444444444444444444444444444444444444444444444444444244444
44444444444444444444444444444444444444444044400444443244444444444444444444
4444444444434 \\
88 & 0.456191 & 0.337710 & 40204444041444424241244444442044444424344022444444444444442444440424044444
44040444044333 \\
101 & 0.511175 & 0.357482 & 44444444444444444444444444444440444444444444444444444344424404444444444244
444442044444444442444444434 \\
59 & 0.532829 & 0.364976 & 44403044442024444434424444444444444444444244404443444042033 \\
44 & 0.585746 & 0.382670 & 44400433344444444444044444444440444444432223 \\
61 & 0.904904 & 0.475633 & 4444444424343424004344444444440444444442444444444230204444403 \\
68 & 0.913233 & 0.477816 & 43444434442444444444440444444444444444444402444404344404442404044423 \\
81 & 0.927386 & 0.481505 & 40444444443344444444434444444444444323444044440244404444434444044442420400
0244234 \\
71 & 1.087120 & 0.521326 & 43434444044044444444244414442441144434404142414444444443143044414420043 \\
56 & 1.108356 & 0.526393 & 44440404444244044344444444240444444244444444134443434421 \\
132 & 1.164335 & 0.539522 & 30401442402020344400402404143444043201444444442420240444404004004042404244
0304040013244240440013044042041040444014012304004400004203 \\
37 & 1.167666 & 0.540293 & 4044404440340444444444444444444430420 \\
39 & 1.193768 & 0.546299 & 434444444444443442444044303404424344443 \\
72 & 1.281908 & 0.566107 & 444240444444444044444444444444434344443444444444044444444322023444444444 \\
169 & 1.288397 & 0.567538 & 02311132200430444041410004402402240444420444404322341404042020004100440004
44240000340114440200344440404044244403414004401444400111004440400403420040
100444444434010044424 \\
47 & 1.313629 & 0.573068 & 44434444344444444444444443133433414444444444433 \\
32 & 1.336065 & 0.577942 & 34444442444432344432244444440433 \\
90 & 1.343007 & 0.579441 & 44444444434444442444042444444014434444444444404444441444041410344404234
0444444344444044333 \\
41 & 1.355551 & 0.582140 & 34434444400234034243044444444002404432303 \\
\bottomrule
\end{tabular}
    \caption{Compilation of CNOT-gate $\alpha=2.4$}
    \label{tab:cnot24_gate_compilation}
\end{table}

\subsection{Local CNOT equivalence gate compilation}

We focus in this experiment on the CNOT local equivalence class, that is, all operations which are equivalent to the CNOT gate up to single qubit transformations from the $SU(2) \otimes SU(2)$ subgroup of $SU(4)$. 
We point out that this significantly enhances the size of the optimization target from a point in fifteen-dimensional space to its twelve-dimensional subspace. This follows from the fact that each element of $SU(2)$ is represented by three parameters, and an arbitrary element of $SU(2) \otimes SU(2)$ is applied before and after a two-qubit gate, such as $\text{CNOT}$, for example, in order to generate the entire local equivalence class. In other words, we can say that, using local invariants, which classify local equivalence classes, we reduce quantum compilation to a three-dimensional problem. We again use the complete alphabet $\text{NON-SEMI}$, including $\text{CPHASE}$ with 
$\alpha$ set to $2.4$. The depth of the circuits is set from $5$ to $80$. 

The results of this experiment are summarized in Table~\ref{tab:cnot_eqclass_gate_compilation}, which shows the best outcomes in descending order. The very best result we were able to obtain comprised of a circuit with depth $35$ and accuracy of $10^{-9}$, as measured by the $D_\text{CNOT}$ distance. This measure is significantly below the typical fault-tolerance threshold of $10^{-4} - 10^{-5}$ (should that ever be needed in this type of systems).
We note that the best outcome is obtained for circuits that almost exclusively consist of the repeated application of the $\text{CPHASE}$ gate from the gate alphabet. 

\begin{table}[]
    \centering
\begin{tabular}{rrl}
\toprule
Depth & Distance & Program \\
\midrule
35 & 1.5617e-09 & 44444444444444444444444444444444444 \\
34 & 1.3831e-05 & 4444444444444444444444444444444444 \\
36 & 2.9260e-05 & 444444444444444444444444444444444444 \\
33 & 2.6959e-04 & 444444444444444444444444444444444 \\
37 & 3.9145e-04 & 4444444444444444444444444444444444444 \\
32 & 1.4467e-03 & 44444444444444444444444444444444 \\
38 & 1.8533e-03 & 44444444444444444444444444444444444444 \\
44 & 2.1372e-03 & 44444444444444444444444434404444444443444444 \\
61 & 2.3334e-03 & 4444444344444444444444444444444434444444443444444444434444444 \\
31 & 4.6776e-03 & 4444444444444444444444444444444 \\
39 & 5.6276e-03 & 444444444444444444444444444444444444444 \\
40 & 5.6277e-03 & 4444444444444444444444444444444444444440 \\
30 & 1.1503e-02 & 444444444444444444444444444444 \\
29 & 2.3804e-02 & 44444444444444444444444444444 \\
41 & 2.6880e-02 & 44444444444444444444444444444444444444444 \\
28 & 4.3733e-02 & 4444444444444444444444444444 \\
42 & 4.8488e-02 & 444444444444444444444444444444444444444444 \\
50 & 6.1524e-02 & 34344444344444444444444444444434444444444404444424 \\
27 & 7.3615e-02 & 444444444444444444444444444 \\
43 & 8.0495e-02 & 4444444444444444444444444444444444444444444 \\
\bottomrule
\end{tabular}

    \caption{Distance to the [CNOT] local equivalence class.}
    \label{tab:cnot_eqclass_gate_compilation}
\end{table}

\subsection{Perfect entangler compilation}
It is reasonable to consider, in addition to local equivalence classes, the complete set of perfect entanglers as possible optimization targets. These are two-qubit operations capable of creating a maximally entangled state out of some initial product, i.e., a non-entangled state. Specifically, we aim to find an arbitrary gate that is a perfect entangler. We use the entire alphabet $\text{NON-SEMI}$, again including $\text{CPHASE}$ and $\alpha$ set to $2.4$. We will keep the depth in the same range as in the previous case, that is, from $5$ to $80$. 

The results are shown in Table~\ref{tab:pe_gate_compilation}, which again shows the best outcomes in descending order with respect to the $D_\text{PE}$ distance. In particular, the very best results for the circuit depth $77$ and $35$ exhibit a distance to the perfect entanglers of zero (in standard 64-bit machine precision). The accuracy of the compiled circuits is improved by several orders of magnitude compared to the same distance metrics of the compilation with respect to the local equivalence classes. We again observe that most of the successful compilations are dominated by the repeated application of the $\text{CPHASE}$ gate. We will address this point in the next section.

\begin{table}[]
    \centering
\begin{tabular}{rrl}
\toprule
Depth & Distance & Program \\
\midrule
77 & 0.0000e+00 & 14044440044022044000440442044024044204420220440440444400040024202404004240221 \\
35 & 0.0000e+00 & 44444444444444444444444444444444444 \\
34 & 3.0622e-11 & 4444444444444444444444444444444444 \\
36 & 1.3698e-10 & 444444444444444444444444444444444444 \\
33 & 8.3225e-09 & 444444444444444444444444444444444 \\
37 & 1.9960e-08 & 4444444444444444444444444444444444444 \\
32 & 3.2197e-07 & 44444444444444444444444444444444 \\
38 & 5.4960e-07 & 44444444444444444444444444444444444444 \\
31 & 3.4437e-06 & 4444444444444444444444444444444 \\
40 & 5.0489e-06 & 4444444444444444444444444444444444444404 \\
39 & 5.0611e-06 & 444444444444444444444444444444444444444 \\
30 & 2.1170e-05 & 444444444444444444444444444444 \\
29 & 9.0532e-05 & 44444444444444444444444444444 \\
41 & 1.1544e-04 & 44444444444444444444444444444444444444444 \\
28 & 3.0602e-04 & 4444444444444444444444444444 \\
42 & 3.7618e-04 & 444444444444444444444444444444444444444444 \\
27 & 8.6702e-04 & 444444444444444444444444444 \\
43 & 1.0365e-03 & 4444444444444444444444444444444444444444444 \\
26 & 2.1474e-03 & 44444444444444444444444444 \\
44 & 2.5122e-03 & 44444444444444444444444444444444444444444444 \\
\bottomrule
\end{tabular}
    \caption{Distance to a perfect entangler.}
    \label{tab:pe_gate_compilation}
\end{table}

\subsection{Gate compilation using only the entangling gate}
As summarized in Tables~\ref{tab:cnot24_gate_compilation}-\ref{tab:pe_gate_compilation}, the MIQCQP quantum compilation found optimal solutions in which the entangling gate CPHASE features prominently. 

This is to be largely expected for the $\text{CPHASE}$ gate given by Eq.~\eqref{eq_entanglingMatrix}. The Mahklin invariants, Eq.~\eqref{eq:g}, $(g_1,g_2,g_3)$ turn out to be close to $(1,0,3)$, which characterizes the local equivalence class of the identity, and hence all single qubit operations from $SU(2) \otimes SU(2)$. In fact, in the representation given by the Cartan decomposition, the local equivalence class [$\text{CPHASE}$] lies on the edge of the Weyl chamber~\cite{ZhangValaBirgitta_GeometricTheory_PRA2003} that connects the local equivalence classes of the identity $[\text{I}]$ and the $[\text{CNOT}]$. Its repeated application then leads to an operation that is closer to the [$\text{CNOT}$] target, eventually reaching it for a sufficiently large number of repetitions.

This is rather remarkable validation of the compilation approach based on the MIQCQP formalism, since the optimal solution it yields agrees with our intuition, which is nevertheless based on sophisticated mathematical formalism (cf. ~\cite{PhysRevA.69.042309}). In particular, Table~\ref{tab:pe_gate_compilation} shows a braiding combination of depth 35, solely consisting of the entangling gate $\text{CPHASE}$, achieving exactly zero distance from the target. In addition to the solution, which clearly captures our intuition, the MIQCQP quantum compilation has also found solutions that would not be accessible in such an intuitive manner.

\begin{figure}[h!]
	\centering 
    \includegraphics[width=0.48\linewidth]{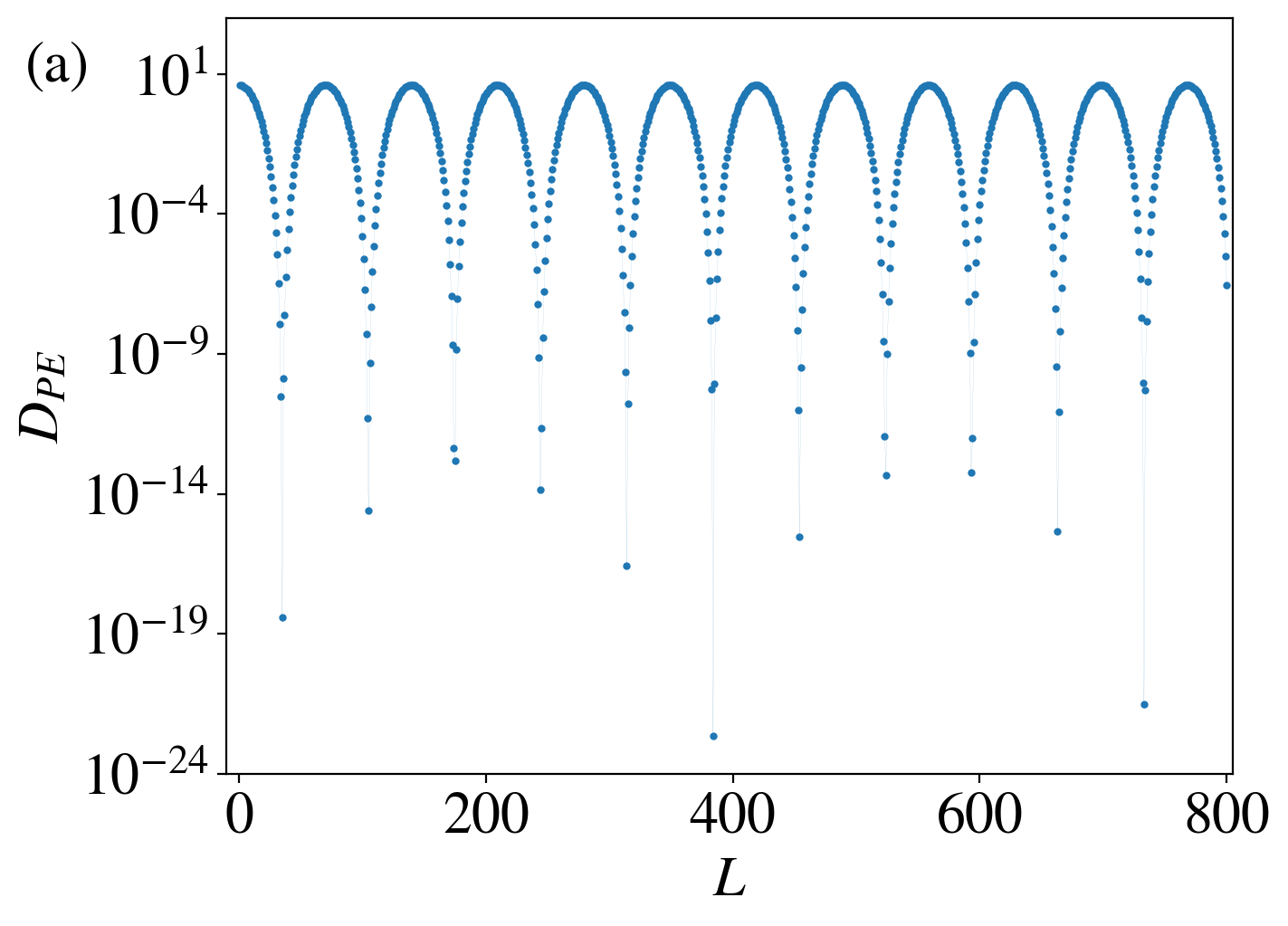} 
    \includegraphics[width=0.50\linewidth]{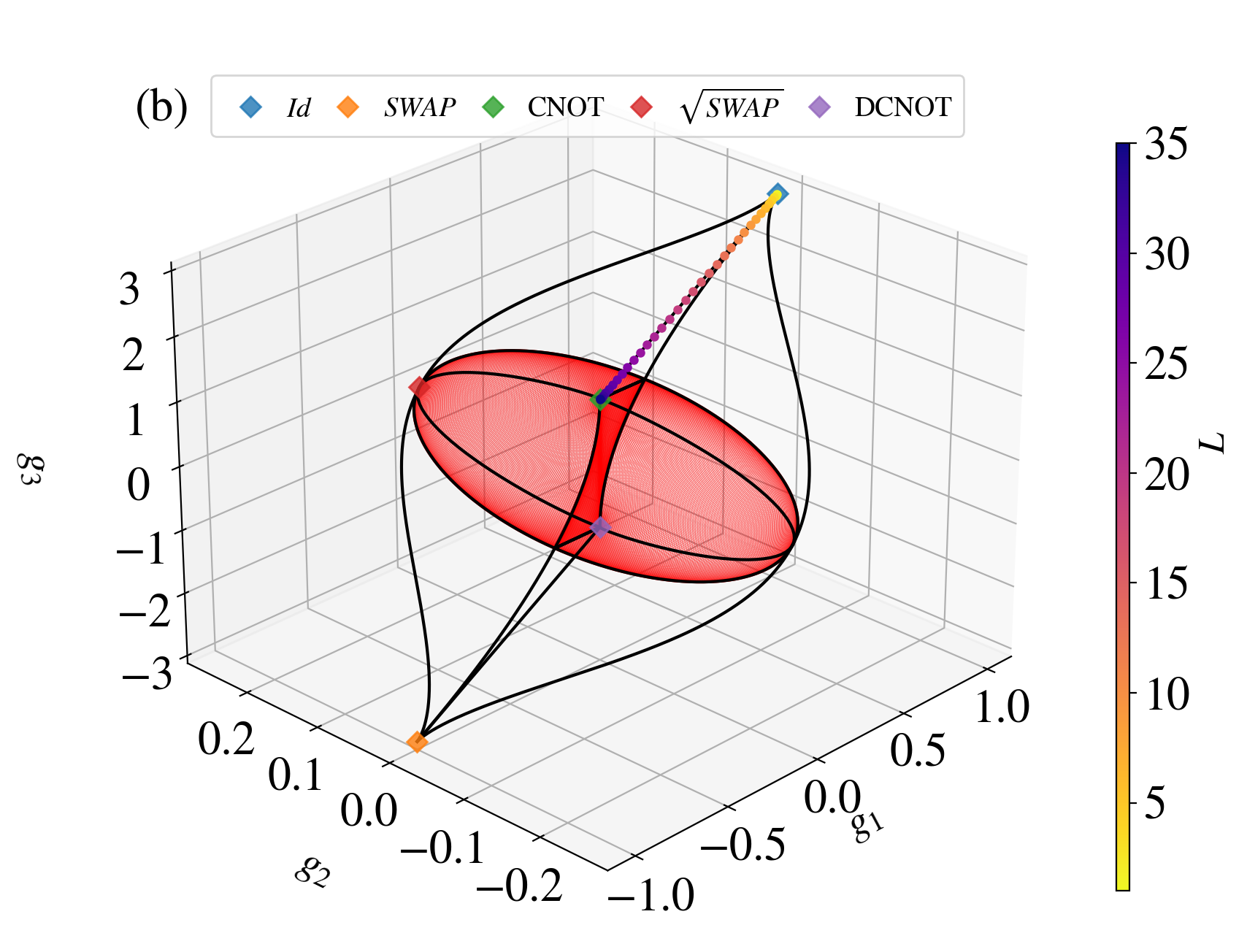} 
	\caption{(a) Distance of powers of the entangling gate (Eq.~\eqref{eq_entanglingMatrix}), to the perfect entanglers (Eq.~\eqref{eq_PE_dist}) plotted versus power $L$. (b) Invariants of matrices from panel (a) plotted in $g_1,g_2,g_3$ space (up to $L=35$). The data is plotted alongside the perfect entangler section of the Weyl chamber, highlighted in red.}
	\label{fig_entangling_dist_vs_depth}
\end{figure}

In order to illustrate the role of $\text{CPHASE}$, we show the distance of successive powers of the entangling gate to the perfect entanglers (Eq.~\eqref{eq_PE_dist}) versus depth or equivalently power $L$ in Figure~\ref{fig_entangling_dist_vs_depth}(a). The data highlights the low distances that can be reached with even moderate depth braiding combinations of solely the entangling gate, with a depth of $\sim35$ achieving a distance which is essentially zero in terms of machine precision. In addition, we plot the invariants of circuits with successive powers of $\text{CPHASE}$ in Figure~\ref{fig_entangling_dist_vs_depth}(b). Here we also illustrate the Weyl chamber in the invariant space, highlighting the perfect entangler region in red. The relevant equivalence classes are highlighted as points on the vertices of the chamber. This plot illustrates the circuit's path along the edge of the Weyl chamber, starting from approximately the identity class and terminating at the CNOT vertex.

\section{Conclusion}\label{sec_conclusions}

We introduced an MIQCQP formulation for quantum circuit compilation. We used McCormick linearization to reduce the number of quadratic constraints. We demonstrated it by compiling the CNOT gate in the non-semi-simple Ising anyons framework. Then we provided an MIQCQP formulation for the distance to a local equivalence class and demonstrated it on compiling a gate that is as close as possible to the local equivalence class of the CNOT gate and also to a perfect entangler. Regarding the perfect entangler in a semi-simple Ising anyon framework, the MIQCQP compilation process arrived at the compilation of the perfect entangler given by a sequence of just one single entangling gate, which is controlled phase rotation and which is native in this anyon system. This is in accord with our intuition regarding controlled phase rotations and the local equivalence classes these operations generate.

\begin{acknowledgments}
P.C.B acknowledges funding from Taighde Éireann - Research Ireland through grant 21/RP-2TF/10019. P.R. and
J.M. acknowledge the support of the Czech Science Foundation (23-0947S).
\end{acknowledgments}
\normalem 
\bibliography{qc_ref}

\end{document}